\documentclass[aps,pop,twocolumn,showpacs,superscriptaddress]{revtex4-1}

\usepackage{amsmath}
\usepackage{graphicx}
\usepackage{color}
\usepackage{bm}
\usepackage{amssymb}

\begin{document}

\title{Scaling Laws of Ion Acceleration in Ultrathin Foils Driven by Laser Radiation Pressure}

\author{X. F. Shen}
\affiliation{Center for Applied Physics and Technology, HEDPS, State Key Laboratory of Nuclear Physics and Technology, and School of Physics, Peking University, Beijing, 100871, China}
\affiliation{Collaborative Innovation Center of IFSA (CICIFSA), Shanghai Jiao Tong University, Shanghai 200240, China}
\author{B. Qiao}
\email[Correspondence should be addressed to: ] {bqiao@pku.edu.cn}
\affiliation{Center for Applied Physics and Technology, HEDPS, State Key Laboratory of Nuclear Physics and Technology, and School of Physics, Peking University, Beijing, 100871, China}
\affiliation{Collaborative Innovation Center of IFSA (CICIFSA), Shanghai Jiao Tong University, Shanghai 200240, China}
\author{H. He}
\affiliation{Department of Physics, Fudan University, Shanghai 200433, China}
\affiliation{Shanghai Institute of Laser Plasma, CAEP, Shanghai, 201800, China}
\author{Y. Xie}
\affiliation{Center for Applied Physics and Technology, HEDPS, State Key Laboratory of Nuclear Physics and Technology, and School of Physics, Peking University, Beijing, 100871, China}
\affiliation{Collaborative Innovation Center of IFSA (CICIFSA), Shanghai Jiao Tong University, Shanghai 200240, China}
\author{H. Zhang}
\affiliation{Center for Applied Physics and Technology, HEDPS, State Key Laboratory of Nuclear Physics and Technology, and School of Physics, Peking University, Beijing, 100871, China}
\affiliation{Institute of Applied Physics and Computational Mathematics, Beijing 100094, China}
\author{C. T. Zhou}
\affiliation{Center for Applied Physics and Technology, HEDPS, State Key Laboratory of Nuclear Physics and Technology, and School of Physics, Peking University, Beijing, 100871, China}
\affiliation{Institute of Applied Physics and Computational Mathematics, Beijing 100094, China}
\author{S. P. Zhu}
\affiliation{Institute of Applied Physics and Computational Mathematics, Beijing 100094, China}
\author{X. T. He}
\affiliation{Center for Applied Physics and Technology, HEDPS, State Key Laboratory of Nuclear Physics and Technology, and School of Physics, Peking University, Beijing, 100871, China}
\affiliation{Collaborative Innovation Center of IFSA (CICIFSA), Shanghai Jiao Tong University, Shanghai 200240, China}
\affiliation{Institute of Applied Physics and Computational Mathematics, Beijing 100094, China}

\date{\today}

\begin{abstract}
Scaling laws of ion acceleration in ultrathin foils driven by radiation pressure of intense laser pulses are investigated by theoretical analysis and two-dimensional particle-in-cell simulations. Considering the instabilities are inevitable during laser plasma interaction, the maximum energy of ions should have two contributions: the bulk acceleration driven by radiation pressure and the sheath acceleration in the moving foil reference induced by hot electrons. A theoretical model is proposed to quantitatively explain the results that the cutoff energy and energy spread are larger than the predictions of ``light sail" model, observed in simulations and experiments for a large range of laser and target parameters. Scaling laws derived from this model and supported by the simulation results are verified by the previous experiments.

\end{abstract}


\maketitle

Laser-driven ion acceleration has the potential to produce compact sources of energetic ions from several MeV to GeV \cite{Macchi2013,Daido2012}, which can be applied for proton radiography \cite{Borghesi2001}, tumor therapy \cite{Bulanov2002a}, inertial fusion energy \cite{Tabak1994} and warm dense matter \cite{Beg2002}. Most of them require ion beams with special energy range and spread. Scaling law studies are devoted to evaluate the laser and target parameters needed to produce ion beams of interest, which has been carried out widely for target normal sheath acceleration (TNSA) \cite{Fuchs2006,Robson2007}, characterized with low scaling ($\propto I^{1/2}$) and broad energy spread \cite{Wilks2001,Schwoerer2006}. However, for the more ideal mechanism radiation pressure acceleration (RPA) \cite{Esirkepov2004,Macchi2005,Robinson2008,Qiao2009,Macchi2009}, only Kar {\it et. al.} \cite{Kar2012} has verified the scaling law of peak energy, while that of maximum energy has never been discussed as we know, though it attracts more interest nowadays. In recent years, several more experiments of RPA have been performed with stronger laser intensity and more advanced target fabrication technology \cite{Henig2009,Bin2015,Scullion2017,Higginson2018}, but the maximum energy of proton beams is still lower than 100MeV. To explain the results and find what conditions should be improved, scaling law studies are necessary and urgently needed.

For the idealized RPA driven by circularly polarized (CP) laser pulses, electron heating should be inhibited and the maximum energy can be described by the ``light sail" (LS) model \cite{Macchi2009}. However, for relatively long, weak and tightly-focused laser pulses, such as pulse duration $\tau_L\sim40\rm{fs}$, laser intensity $I_0\sim10^{20}\rm{W/cm^2}$ and spot $r\sim2\mu m$ (the commonly used parameter range in nowadays experiments \cite{Henig2009,Bin2015,Scullion2017}), various instabilities, including transverse instabilities \cite{Pegoraro2007} and finite spot effects \cite{Dollar2012}, have enough time to set in and grow up to the nonlinear phase. The foil surface will be deformed and laser is no longer normally incident, leading to serious electron heating. Just like in TNSA, the hot electrons can form a strong charge-separation field at the target rear surface, which can accelerate ions and broaden the energy spectrum even after the pulse ends. Thus the cutoff energy is much higher and the energy spread larger than the predictions of LS model, which has been observed in the previous experiments \cite{Kar2012,Henig2009,Bin2015,Scullion2017,Higginson2018}. The detrimental effects of hot electron generation to the acceleration have been investigated extensively \cite{Robinson2008,Shen2017}, however, its contribution to the scaling laws of RPA has not been discussed adequately.

In this paper, we investigate the scaling laws of the maximum ion energy for ion acceleration in ultrathin foils driven by laser radiation pressure. As plenty of hot electrons generate due to instabilities during the laser plasma interaction, the acceleration includes not only the bulk acceleration driven by radiation pressure and coinciding with LS model \cite{Macchi2009}, but also the sheath acceleration in the foil reference frame, caused by hot electrons and described by the two-phase model including adiabatic electron cooling \cite{Mora2005}. The maximum energy in the laboratory frame can be obtained through the Lorentz transformation, which can quantitatively explain the simulation results for a large range of laser and plasma parameters. The scaling law is given by fitting the maximum ion energy obtained from simulations as a function of laser and target parameters, which also accords with the previous experiment results very well.

When an intense CP laser irradiates an ultrathin opaque foil, the acceleration of the bulk target is dominated by LS RPA, the peak energy per nucleon of which is given by \cite{Macchi2009}
\begin{equation}
\epsilon_{p}=\frac{\xi^2}{2(1+\xi)}m_{p}c^2,\\
\label{eq:ei}
\end{equation}
where $\xi=2\pi\frac{Z}{A}\frac{m_{e}}{m_{p}}\frac{a_0^2\tau}{\zeta}$, $Z/A$ is the charge to mass ratio, $m_e$ and $m_p$ are the electron and proton mass, $\zeta=\pi\frac{n_e}{n_c}\frac{l}{\lambda}$, $a_0$, $\tau$, $n_c$, $n_e$, $l$ and $\lambda$ are the normalized laser intensity, the normalized pulse duration, the cutoff density, the initial electron density, the target thickness and the laser wavelength, respectively.

As the bulk foil is pushed forward by the laser pulse, instabilities, such as Rayleigh-Taylor-like instability \cite{Pegoraro2007} and finite spot effects \cite{Dollar2012}, set in and increase during the laser-plasma interactions, which lead to serious electron heating. Hot electrons generate and build a charge separation field (sheath field) at the target rear side, which could accelerate the ions from the outermost of the foil to much higher energy than the prediction of LS model. In the foil reference frame, the acceleration progress is just like that of a thin-foil expansion into vacuum described by Mora \cite{Mora2005}, which gives:
\begin{equation}
\epsilon_{T}=2\frac{Z}{A}\alpha T_e,
\label{eq:et}
\end{equation}
where $T_e$ is the electron temperature, $\alpha=[{\rm{ln}}(0.32l/\lambda_d+4.2)]^2$ in the Ref. \cite{Mora2005} with Debye length $\lambda_d=\sqrt{T_e/4\pi n_ee^2}$, which should be modified according to the specific parameters of laser and target \cite{Robson2007}. This acceleration progress is actually ubiquitous for RPA considering that instabilities are inevitable, but has never been discussed in detail.

In the laboratory frame, the maximum ion energy obtained from a Lorentz transformation is:
\begin{equation}
\epsilon_{max}=\frac{\epsilon_{p}\epsilon_{T}}{m_ic^2}[1+\sqrt{1+\frac{2m_ic^2}{\epsilon_{p}}}\sqrt{1+\frac{2m_ic^2}{\epsilon_{T}}}]+\epsilon_{p}+\epsilon_{T}.\\
\label{eq:emax}
\end{equation}
Actually, in the foil reference frame, the sheath field exists both at the front and rear of the foil. The sheath field at the rear accelerates ion forward, while the front accelerates that backward, which, in fact, decelerates ions when we change the frame to the laboratory frame, corresponding to $\epsilon_{min}$. Thus, we suggest that the energy spread can be estimated by $\Delta=(\epsilon_{max}-\epsilon_{min})/2\epsilon_p$, then we have $\Delta=\sqrt{1+2\frac{m_ic^2}{\epsilon_p}}\sqrt{\frac{\epsilon_T}{m_ic^2}(\frac{\epsilon_T}{m_ic^2}+2)}$. In weak relativistic limit, the maximum energy and energy spread can be approched as:
\begin{eqnarray}
\epsilon_{max}\approx(\sqrt{\epsilon_{p}}+\sqrt{\epsilon_{T}})^2,\\
\label{eq:em2}
\Delta\approx2\sqrt{\frac{\epsilon_T}{\epsilon_p}}.\;\;\;\;\;\;\;\;\;\;\;\;\;\;\;\;
\label{eq:de}
\end{eqnarray}

To elucidate the dynamics we described above, 2D PIC simulations are carried out with the EPOCH code \cite{Arber2015}. The simulation box ($x,y$) is 14.4$\mu\rm{m}\times$24$\mu\rm{m}$ containing 14400$\times$4000 cells. The foil is a purely hydrogen plasma with density $200n_c$ and thickness 14nm, given by the optimal thickness $l_o=a_0n_c\lambda/\pi n_e$ of LS model. The particle number per cell for electrons and ions is 200. A CP laser pulse with $a_0=10$, $\lambda=800$nm is normally incident on the foil which locates at $x=0$. And to better compare with the above theory, the laser pulse has a transversely fourth-order Gaussian profile with spot radius $r=5\mu m$ and temporally flattop envelope (1$T_0$ rise and fall times and 14$T_0$ plateau). In the following, the pulse duration are calculated as $\tau=t_p+1T_0$, where $t_p$ is the length of plateau. To show the effect of foil thickness, a simulation with thickness 22nm and other parameter unchanged is carried out for comparison.

\begin{figure}
\includegraphics[width=8.6cm]{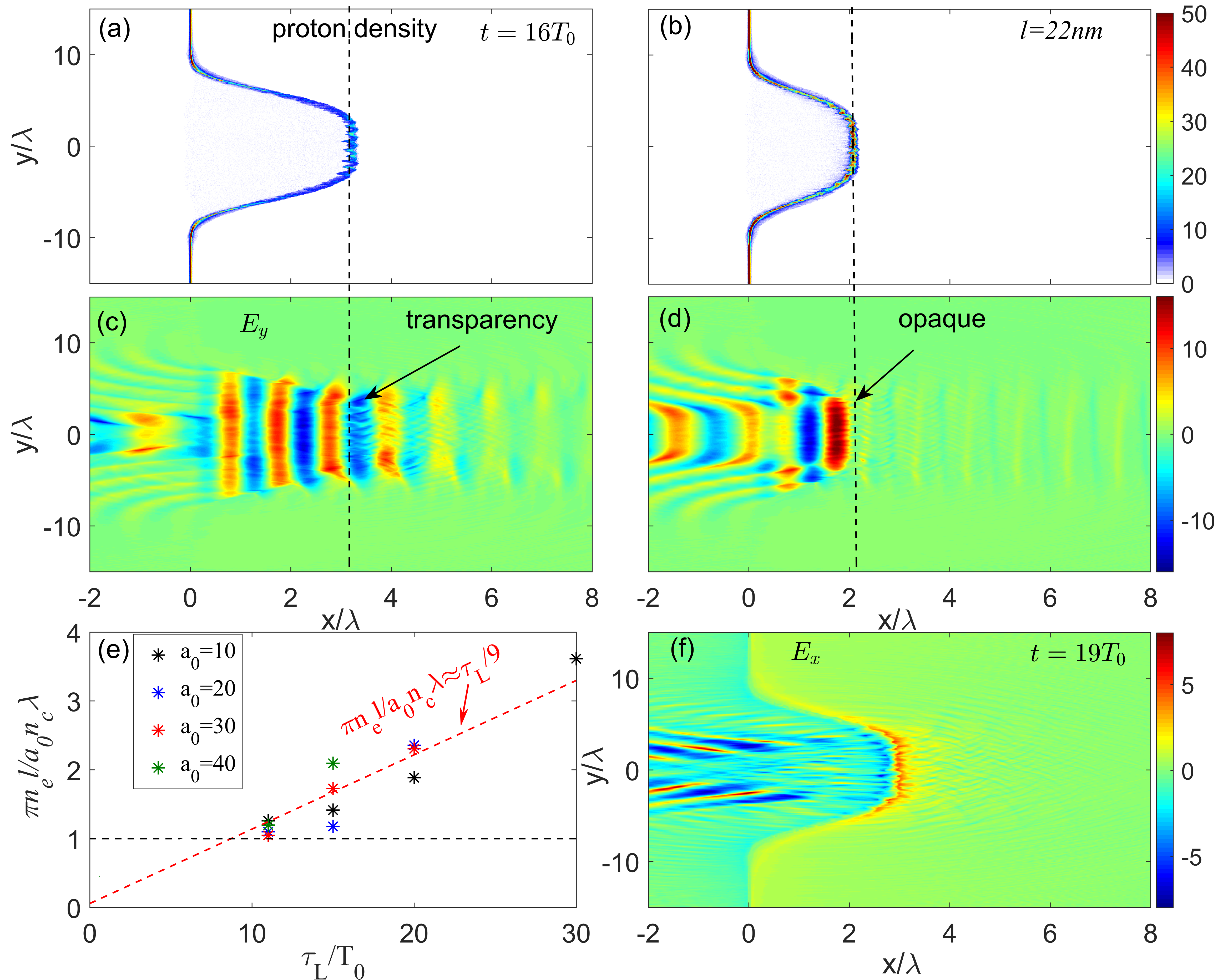}
\caption{(color online) Proton density and laser field ($E_y$) distribution at $t=16T_0$ for the target thickness 14nm [(a) and (c)] and 22nm [(b) and (d)], respectively, where $a_0=10$, $\tau_L=15T_0$ and $n_e=200n_c$. (e) the thickness versus the pulse duration to keep the foil opaque until the pulse ends. The black, blue, red and green stars correspond to $a_0=10$, 20, 30 and 40. The dashed red line shows the tendency, while the dashed black line represents the classical condition. (f) the longitudinal electric field at $t=19T_0$, corresponding to the time when the laser pulse just ends.} \label{fig:figftwo}
\end{figure}

\begin{figure}
\includegraphics[width=8.0cm]{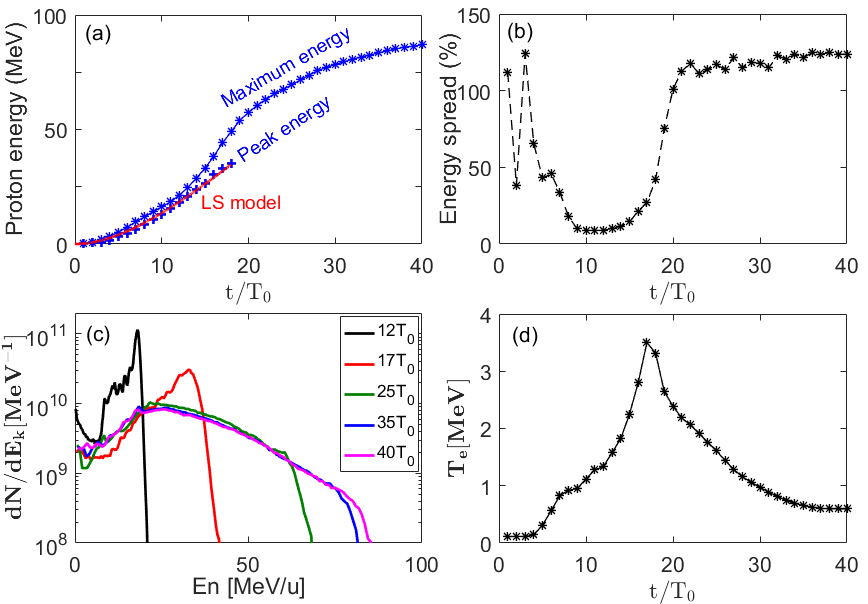}
\caption{(color online) The evolution of proton energy (a), energy spread (b), energy spectra (c) and electron temperature (d) for the case with $l_0=22$nm. In (a), the dashed blue line and the blue crosses show the maximum and peak energy of ion beams varying with time from simulation, respectively, while the red line represents the prediction of LS model. In (b) and (c), the energy spread and spectra of ion beams are counted within $\mid y \mid<4\mu m$ and angular $\theta<10^\circ$. In (d), the electron temperature is represented by the mean electron energy within the laser spot.} \label{fig:figone}
\end{figure}

Figure \ref{fig:figftwo}(a) and (c) show the distributions of proton density and laser field at $t=16T_0$ for the foil with thickness 14nm, respectively. The foil is thick enough to keep the balance between electrostatic pressure and radiation pressure at the beginning. However, as the instabilities set in and increase with time, plenty of superthermal electrons generate and escape. Then the electrostatic pressure decreases and is not large enough to keep balance to the radiation pressure. Meanwhile, the relativistic induced transparency (RIT) \cite{Vshivkov1998} could also occur due to plasma expansion and relativistic increase of electron mass caused by electron heating. Finally, the pulse punches through the foil [Fig. \ref{fig:figftwo}(c)] and the bulk acceleration ends prematurely, which means the optimal thickness $l_o$ suggested by the LS model is not large enough to keep the bulk acceleration lasts till the pulse ends for the relatively long and weak laser pulses. However, this can be achieved if we use a thicker foil, as shown in Fig. \ref{fig:figftwo}(b) and (d). As the growth rate of instabilities decreases with increase of the foil thickness \cite{Pegoraro2007}, which means less hot electron loss and smaller electron energy. Thus not only the electrostatic pressure is large enough to balance the radiation pressure, but also the RIT is suppressed. In the following, we note ``stable RPA" for the cases that bulk acceleration maintains until the pulse ends, while ``unstable RPA" for the cases not.

To better understand the acceleration progress, figure \ref{fig:figone} shows the evolution of the proton energy [\ref{fig:figone}(a)], energy spread [\ref{fig:figone}(b)], energy spectra [\ref{fig:figone}(c)] and electron temperature [\ref{fig:figone}(d)] for the case with $l_0=22$nm.
The foil keeps opaque until the pulse ends at $t=17T_0$, where we can see a turning point in Fig. \ref{fig:figone}(a) and (d), before which the acceleration is mainly driven by radiation pressure. The bulk foil is accelerated forward and a peak near the cutoff energy in the energy spectrum is still very clear at $t=17T_0$ [the red line in \ref{fig:figone}(c)]. After that the acceleration is dominated by sheath acceleration, which accelerates ions from the outermost and the faster ions experience stronger acceleration field, as shown in Fig. \ref{fig:figone}(c). Thus the ion energy increases [\ref{fig:figone}(c)] and so does the energy spread [\ref{fig:figone}(b)]. At the end of the acceleration $t=40T_0$, the energy spread and electron temperature keeps almost unchanged and the increase of the maximum energy is very small, which means the increase of the cutoff energy is less than 1$\%$ in the next pulse cycle. The final proton energy is about $84$MeV, which is much larger than the peak energy $32$MeV predicted by the LS model and the maximum energy $44$MeV when the pulse ends. It should be noted that the peak energy evolution still can be described by the LS model [the blue crosses and red line in \ref{fig:figone}(a)]. The energy spread is about $124\%$, exponentially decreasing in the high energy [Fig. \ref{fig:figone}(c)].

During the bulk acceleration, the electron temperature increases due to laser-plasma interaction [\ref{fig:figone}(d)]. It only keeps relatively low before $t=12T_0$, then increases quickly, as the target is seriously deformed (shown in Fig. \ref{fig:figftwo}(b) and (d)) and the highest electron temperature is about $3.5$MeV at $t=17T_0$. Here we use the mean electron energy within the laser spot to represent the temperature. After the pulse ends, electron temperature decreases slowly, which means the energy of electrons transfers to that of protons, just like the description of the two-phase model \cite{Mora2005}. Moreover, figure \ref{fig:figftwo}(f) shows the longitudinal electric field at $t=19T_0$. According to the two-phase model \cite{Mora2005}, $E_{front}=\sqrt{2/e_{\rm N}}T_e/e\lambda_d\approx5.8E_0$, where $e_{\rm N}\approx2.71828$ and $E_0=m_e\omega c/e$, which is slight higher than the simulation results [Fig. \ref{fig:figftwo}(f)]. Though the acceleration of a positive plate can also be described by the model of Coulomb explosion \cite{Landau1988,Bulanov2002b}, it is very complicated considering the electron oscillation and hard to estimate the longitudinal electric field exactly \cite{Fourkal2005,Grech2011}. Thus we think the two-phase model is more suitable to describe the acceleration progress after the pulse ends.

To quantitatively prove that Eq. 4 and 5 can estimate the maximum energy and energy spread obtained from simulations, we still need to know the parameter $\alpha$. As we hope it can be applied to a large range of laser and plasma parameters, not just the case we discussed above. Thus we have performed a series of simulations: the intensity varies from $I=10^{20}$ ($a_0=5$) to $1.5\times 10^{22}$ ($a_0=60$) W/cm$^2$ (7 samples); the pulse duration changes from $\tau=8T_0$ to $30T_0$ (5 samples), while the target thickness is related to the laser intensity and pulse duration. The laser profiles and target density keep the same as above. Here, we focus on the stable RPA for simplicity. Otherwise, we need to know the time when transparency happens, which is not so convenient, especially in experiments. The condition to achieve this is shown by the asterisks as a function of $\zeta/a_0$ and $\tau$ in Fig. \ref{fig:figftwo}(e). The dashed red line is the fitting result, which suggests:
\begin{equation}
\zeta\approx\frac{\tau}{9}a_0.
\label{eq:mc}
\end{equation}
Thus to achieve stable RPA, we need thicker foils for longer pulses, as the instabilities increase with time, but decrease with foil thickness. With the condition Eq. \ref{eq:mc}, the instabilities could be kept at a reasonable level and stable RPA achieved. Meanwhile, the conversion efficiency of laser-to-ion is the highest, as the reflectivity $R\simeq1$ during the bulk acceleration. 

\begin{figure}
\includegraphics[width=8.0cm]{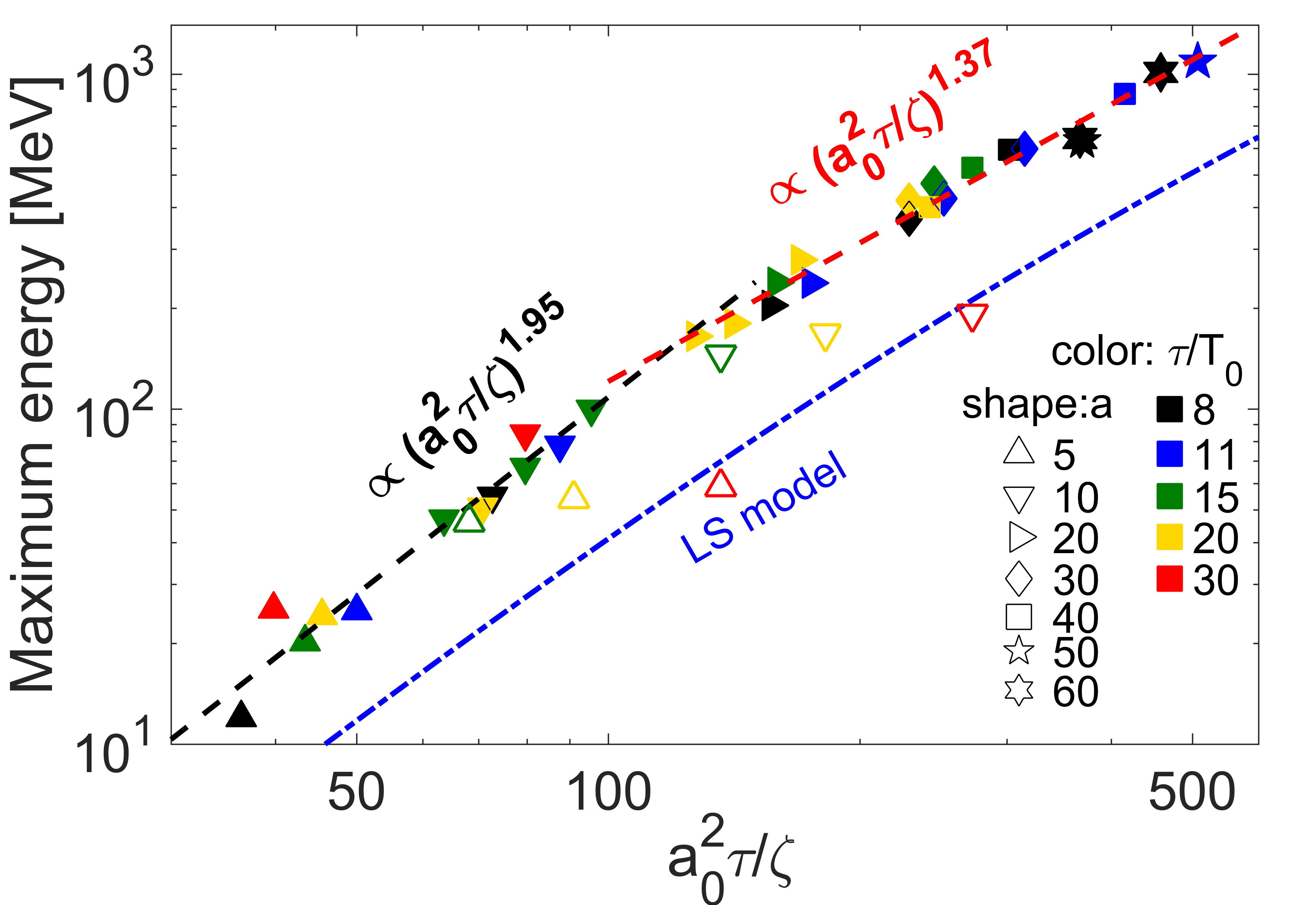}
\caption{(color online) Scaling of the maximum proton energy with $a_0^2\tau/\zeta$. The filled symbols represent the stable RPA and the open ones show the unstable RPA. The colors stand for the pulse duration and the shapes correspond to the laser intensity. The dashed black and red line show the best fit curves for ion energy less and larger than $150$MeV, respectively. The dashed-dotted blue line represents peak energy obtained from LS model.} \label{fig:figtwo}
\end{figure}

Then we plot the correlations of the maximum energy with the parameter $a_0^2\tau/\zeta$, as shown in Fig. \ref{fig:figtwo}. The value of $\alpha=1.25$ is confirmed as the one fitting the simulation results best. With this, the sum of the absolute errors between the energy given by Eq. 4 and simulations is the minimum. In other words, the error is always less than $10\%$ for a large range of laser and plasma parameters, while the results from Eq. \ref{eq:ei} always underestimate that by larger than $50\%$, which means the contribution of sheath acceleration is always up to about half of the final energy and the energy spread is very large, rather than small as the expectation of LS model. Then for the case $l_0=22$nm, we have $\epsilon_T\approx8.8$MeV, thus the maximum proton energy can be calculated from Eq. \ref{eq:emax} or Eq. 4, which is about $75$MeV. And the energy spread estimated by Eq. \ref{eq:de} is about $106\%$. Both of them are very close to those of the simulation results. Thus we have quantitatively verified that our theoretical model is able to describe the maximum energy and energy spread of ion beams obtained from RPA. Besides, though $\alpha$ is obtained from the fit of stable RPA, it also can be used to estimate the unstable cases, if we know the penetration time. Taking the case $l_o=14nm$ discussed before, the penetration time is around $t=15T_0$, thus the effective pulse length for bulk acceleration is about $\tau^{'}=12T_0$. With $T_e^{'}=8.4$MeV, Eq. 4 gives the maximum energy $\epsilon_{max}^{'}=135$MeV, comparing with $154$MeV from simulation.

\begin{figure}
\includegraphics[width=8.0cm]{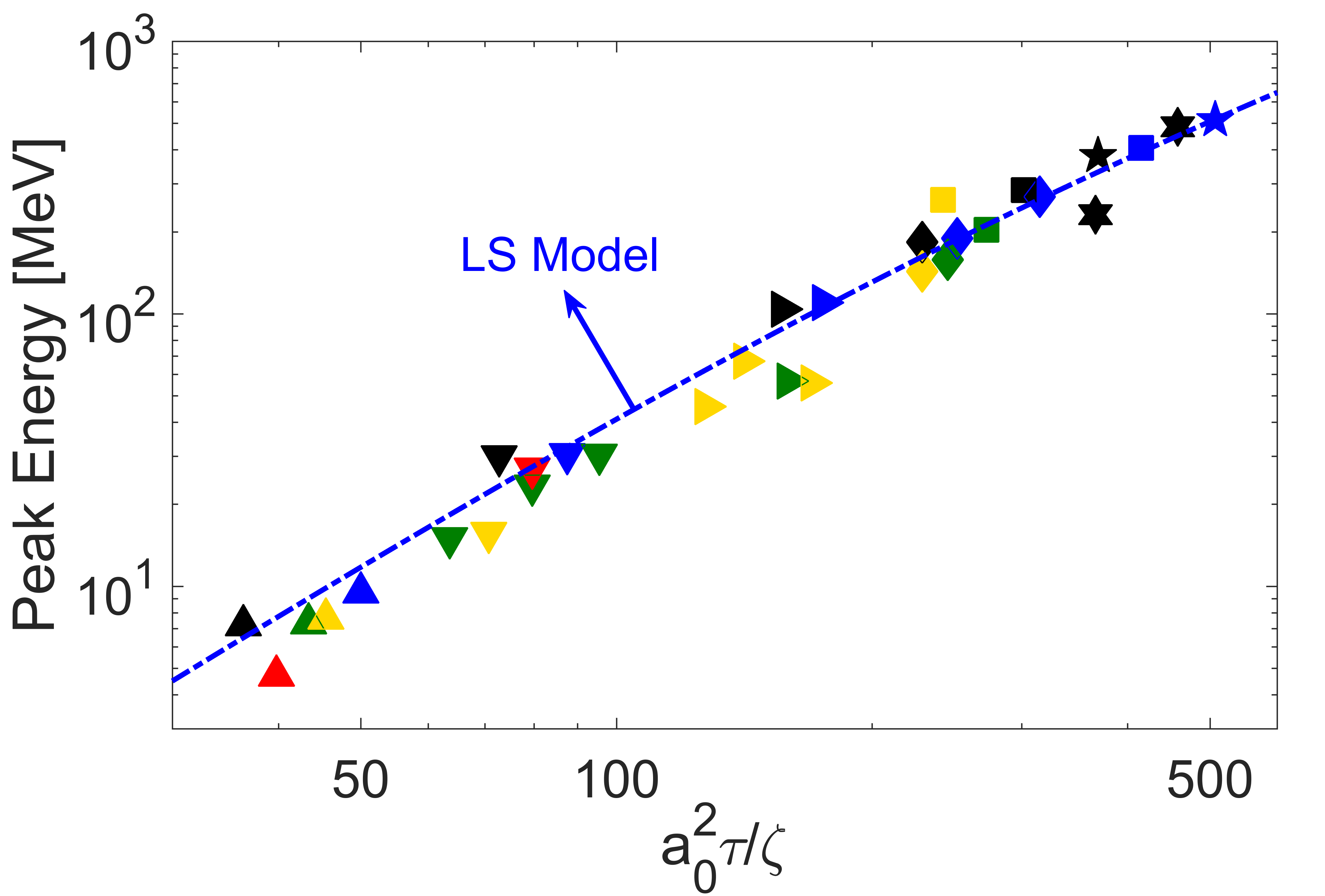}
\caption{(color online) Scaling of the peak proton energy with $a_0^2\tau/\zeta$. The symbols and line represent the same meaning of Fig. \ref{fig:figtwo}.} \label{fig:figpeak}
\end{figure}
However, even we know the parameter $\alpha$, Eq. \ref{eq:emax} is still not convenient for evaluating the maximum energy only from the laser and target parameters, as the quantitatively relationship between $T_e$ and laser/target parameters is quite complicated and still not clear. Thus to obtain a scaling law which is suitable to use, we try to find the best fit between $\epsilon_{max}$ and $a_0^2\tau/\zeta$, just depending on the laser and target parameters. For $\xi<<1$, shown by the dashed black line in Fig. \ref{fig:figtwo}, we have:
\begin{equation}
\epsilon_{max}\approx0.0135\frac{Z}{A}(a_0^2\tau/\zeta)^{1.95} [{\rm MeV/\mu}].
\label{eq:scaling}
\end{equation}
It is interesting to note that the scaling law of the ion maximum energy Eq. \ref{eq:scaling} still indicates that with the same $a_0^2\tau/\zeta$, the maximum ion energy is similar, just like that of peak energy suggested by LS model and demonstrated in Ref. \cite{Kar2012}. However, the parameter (0.0135) in Eq. \ref{eq:scaling} is much larger than that (0.0055) of LS model (the dashed-dotted blue line in Fig. \ref{fig:figtwo}). In addition, we show the scaling of the peak energy with $a_0^2\tau/\zeta$ in Fig. \ref{fig:figpeak}, which still accords with the LS model (Eq. 1) though the further acceleration may affect the peak energy.  Note with Eq. 4, the other parameter $Z/A$ should be more complicated and still relies on $T_e$, so we just simplify it as $Z/A$, though it may slightly overestimate the maximum energy for high-Z ions. If we substitute Eq. \ref{eq:mc} into Eq. \ref{eq:scaling}, we have the scaling law only related to laser intensity: $\epsilon_{max}\approx 0.98Z/Aa_0^{1.95}$ [MeV$/\mu$]. For $a_0=10$, the upper limit is only about 90MeV, just as the filled downward-pointing triangles shown.

Moreover, we can see the maximum energies of the open triangles in Fig. \ref{fig:figtwo} are higher than those of filled, but tend to saturate for long pulse duration, where the foil thicknesses keep unchanged for the same laser intensity. This should be owe to the contributions of sheath acceleration because of higher electron temperature for longer pulse, while the bulk acceleration has reached saturation when the transparency happens. Substituting Eq. \ref{eq:mc} into Eq. 4, after some algebra derivations, we have $\epsilon_{max}\propto(I_0^{1/2}+I_0^{1/4}\tau^{1/2})^2$, where $T_e\approx\eta I_0\tau/n_el\propto\eta I_0^{1/2}\tau$ is estimated by equating the plasma electron energy density to the absorbed laser energy density \cite{Fiuza2012}, with $\eta$ the absorption efficiency,
which means the maximum energy is weakly related to the pulse duration and will saturate for long pulses [Fig. \ref{fig:figtwo}], as $T_e$ has an upper limit. If we assume that the upper limit of $T_e$ is given by the pondermotive scaling $T_{max}=0.511(\sqrt{a_0^2+1}-1)$ MeV \cite{Wilks1992}. Then we have:
\begin{equation}
\epsilon_{max}\approx (0.67\frac{Z}{A}a_0+1.6\sqrt{\frac{Z}{A}a_0})^2 [{\rm MeV/\mu}].
\label{eq:ema}
\end{equation}
For $a_0=5$ and 10, Eq. \ref{eq:ema} predicts the maximum energies are about $48$MeV and $150$MeV, respectively, which are very close to the upper limits of the tendencies shown by the open upward and downward pointing triangles in Fig. \ref{fig:figtwo}, respectively. Note that though the laser intensities used in Refs. \cite{Kar2012,Bin2015,Scullion2017,Higginson2018} are already slight larger than $a_0=10$, the challenge of proton energy larger 100MeV is still not achieved. The main reason is that the bulk speed of the foil is low, which is mainly caused by the high-Z material targets, as the acceleration of that is much smaller than that of protons due to small $Z/A$ (Eq. \ref{eq:ei}). On the other hand, laser pulses with Gaussian temporal profile become another limitation if the transparency happens before the peak of the laser pulse. Besides, we have the energy spread $\Delta\propto\sqrt{\tau/I_0}$ by substituting Eq. \ref{eq:mc} into Eq. 5, which means larger energy spread for longer laser pulses.

For $\xi>>1$, the scaling suggested by LS model becomes unfavorable as $\epsilon_p\propto I^{1/2}$, which is within our expect. However, in our simulations, shown by the dashed red line in Fig. \ref{fig:figtwo}, the scaling law degrades to $\epsilon_{max}\approx 0.2163(a_0^2\tau/\zeta)^{1.37}\propto I^{0.68}$ quickly, just like that of TNSA. The transition point is only about $200$MeV. However, we still think that RPA is the more efficient acceleration mechanism compared with TNSA, as the parameter is higher and so is the conversion efficiency of laser-to-ion \cite{Fuchs2006}.

\begin{figure}
\includegraphics[width=7.6cm]{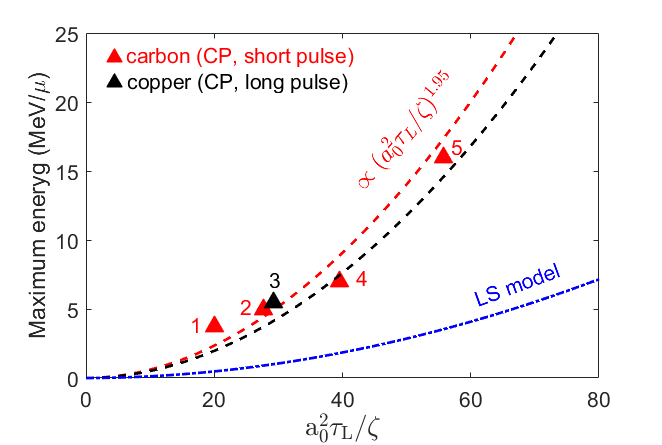}
\caption{(color online) Comparison the experiment data with the scaling law (Eq. \ref{eq:scaling}). The red triangles and dashed red line represent the experiment data and scaling law for carbon ions (C$^{6+}$), while the black for copper ions (Cu$^{27+}$). The data points are obtained from Refs. \cite{Henig2009} (point 1), \cite{Scullion2017} (point 2 and 5), \cite{Kar2012} (point 3) and \cite{Bin2015} (point 4).} \label{fig:fignfour}
\end{figure}

To confirm the scaling law Eq. \ref{eq:scaling} is not artificial of simulation results, we compare the maximum carbon ion energies calculated by Eq. \ref{eq:scaling} with those observed in experiments previously published in Refs. \cite{Henig2009,Kar2012,Bin2015,Scullion2017}. Figure \ref{fig:fignfour} shows the comparison, where the dashed red and black lines show the scaling law of Eq. \ref{eq:scaling} for carbon and copper ions, which correspond to the materials of the foil used in experiments, and the triangles represent the experiment results. Meanwhile the dashed-dotted blue line stands for the prediction of the LS model. It is clear that the ion energies obtained from experiments are significantly higher than that calculated by the LS model, but are quantitatively according with our scaling law Eq. \ref{eq:scaling}. The reason that the scaling law obtained from 2D simulations is able to predict the real 3D experiments is because for stable RPA or transparent not so early, the differences of the maximum energy between 2D and 3D simulations are very small, which has been pointed out in Refs. \cite{Shen2017,Scullion2017}. This enables us to evaluate the laser and target parameters needed to produce high energy ion beams of interest. And this is the first scaling law, to our knowledge, which can quantitatively explain the maximum ion energies obtained from different experiment results of stable RPA (which actually covers the main experiment results published before).

In conclusion, we propose the scaling laws with theoretical analysis and 2D PIC simulations and supported by the present experiment results. The theoretical model is based on the cascade function of bulk acceleration driven by radiation pressure and sheath acceleration induced by hot electrons, which gives a reasonable explanation about the higher cutoff energy and larger energy spread than the predictions of LS model observed in experiments and simulations. Meanwhile, the scaling law shows that the maximum ion energy is mainly limited by the laser intensity. To achieve proton energy larger than 100MeV, we hope the laser intensity is larger than $2.0\times10^{21}{\rm W/cm^2}$, which would be soon attained in the laser device with $5\rm{PW}$ \cite{Chu2015}, and the acceleration scheme should be the unstable RPA or the hybrid RPA-TNSA proposed by Qiao $et.$ $al.$ \cite{Qiao2012} and demonstrated by Higginson $et.$ $al.$ \cite{Higginson2018}, recently.

\end{document}